\documentstyle[preprint,aps,prl]{revtex}
\input psfig
\begin{document}
\draft
\preprint{FNAL-Pub-94/114-A}
\date{May 1994}
\title{
Microwave background anisotropy in low-$\Omega_0$
inflationary models\\ and
the scale of homogeneity in the Universe}

\author{A. Kashlinsky$^{1}$, I. I.\ Tkachev$^{2,3}$ and J. Frieman$^{2,4}$}
\address{
$^{1}$Code 685, Goddard Space Flight Center, Greenbelt, MD 20771}
\address{
$^{2}$NASA/Fermilab Astrophysics Center\\
Fermi National Accelerator Laboratory, Batavia, IL~~60510}
\address{
$^{3}$Institute for Nuclear Research
Russian Academy of Sciences, Moscow 117312, Russia}
\address{
$^{4}$Department of Astronomy and Astrophysics, University of Chicago,
Chicago, IL 60637}

\maketitle

\begin{abstract}
We study the microwave background anisotropy due to superhorizon-size
perturbations (the Grischuk-Zel'dovich effect) in open universes with
negative spatial curvature. Using  COBE results on the
low-order temperature multipole moments, we find that if the homogeneity
of the observable Universe arises from an early epoch of
inflation, the present density parameter cannot differ
from unity by more than the observed quadrupole anisotropy, $|1-\Omega_0|
\alt Q \simeq 5\times 10^{-6}$. Thus, inflation models with low $\Omega_0$
either do not fit
the microwave background observations or they do not solve the horizon problem.
\end{abstract}

\narrowtext

The inflationary scenario for the very early universe has
proven very attractive, because it can simultaneously
solve a number of cosmological puzzles, such as the
homogeneity of the Universe on scales exceeding the particle horizon
at early times, the flatness or entropy problem, and the origin of
density fluctuations for large-scale structure \cite{ko90}. In this
scenario, the
observed Universe (roughly, the present Hubble volume)
represents part of a homogeneous inflated region embedded
in an inhomogeneous space-time. On scales beyond the
size of this homogeneous patch, the initially inhomogeneous
distribution of energy-momentum that existed prior to inflation is preserved,
the scale of the inhomogeneities merely being stretched by the expansion.

In its conventional form, inflation predicts a nearly scale-invariant
spectrum of
density perturbations produced by the inflaton field, and that the
Universe is observationally indistinguishable from being spatially flat
($k=0$). In the absence of a cosmological constant or exotic forms of
matter, this implies that the present matter
density parameter $\Omega_0 \equiv 8\pi G \rho_m(t_0)/3H^2_0$ is very
close to unity. However, it is not
clear that such an Einstein-de Sitter Universe jibes with astronomical
observations.
As is well known, dynamical estimates of mass-to-light ratios from
galaxy rotation curves and cluster dynamics \cite{kg82} typically indicate
$\Omega_0 \simeq 0.1 - 0.2$. Similar conclusions have recently been reached
from the consistency of the ROSAT observations of X-ray emission
from the Coma cluster and Big Bang nucleosynthesis constraints on
the baryon density $\Omega_B$ \cite{wnef}. Determination of the density
parameter using
least-action tracing of the Local Group orbits back in time also
points to a low value of $\Omega_0$ \cite{p89}. Moreover,
if $\Omega_0 = 1$  the
age of the Universe is $t_0 =(2/3H_0) = 6.7 \times
10^9h^{-1}$yrs (where
the present Hubble parameter is $H_0 = 100h$ km/sec/Mpc).
This is less than globular cluster age estimates of
$t_{gc} \simeq 13 - 15 \times 10^9$ yr if $h \geq 0.5$, and a
number of extragalactic distance indicators suggest $h \simeq 0.8$.
A large age is also indicated by the colors of stellar populations of radio
galaxies at high redshift, $z\simeq4$ \cite{cc90}. The presence of
galaxies and perhaps even protoclusters at $z\geq 3.5$ is also easier to
explain in a low-density Universe, where structures should have collapsed by
$z\simeq \Omega_0^{-1} -1$ \cite{gk}. On larger scales, the
situation is still uncertain: several analyses of large-scale peculiar motions
suggest higher values of
$\Omega_0$, consistent with unity \cite{dekrev}, while
other methods are consistent with low values of $\Omega_0$ \cite{d93}.
Finally, the hot and cold spots in the
microwave background recently found by the Tenerife experiment
could perhaps be an imprint of the curvature scale in an open universe
\cite{h94}.

In sum, the current observational status of
$\Omega_0$ is at best inconclusive, with much of the data
pointing to a low-density Universe. In the context of inflation,
the simplest way to accomodate $\Omega_0 < 1$ is to incorporate
a cosmological constant $\Lambda=3H^2_0 \Omega_\Lambda$, retaining spatial
flatness by imposing $\Omega_0 + \Omega_\Lambda = 1$.
However, initial studies of observed
gravitational lens statistics indicate the bound $\Omega_\Lambda \alt
0.7$ \cite{fp}, marginally disfavoring the spatially flat, low-density model.

The other logical possibility is an open, negatively curved universe, and
various suggestions have been made to try to accommodate an open,
low-$\Omega_0$ Universe within inflation \cite{loi}. While the models
differ in the
mechanisms that drive inflation, their common feature is that the
homogeneous patch that
encompasses the presently observable Universe was inflated by just the
right number
of $e$-foldings to ensure that $1-\Omega_0\simeq 1$; generally, this
implies that the present size $L_0$ of the
inflated patch is comparable to the current Hubble distance, $H^{-1}_0$.

Points separated by distances larger than the scale
of the inflated homogeneous patch have never
been in causal contact, and one thus expects large density fluctuations,
$(\delta \rho/\rho)_L \sim 1$, on scales $L \agt L_0$.
However, if the size of the homogeneous region is close to the
present Hubble radius, such non-linear inhomogeneities on large scales
will induce significant microwave background anisotropy
via the Grischuk-Zel'dovich (GZ)  effect \cite{gz78}. In order
of magnitude, the quadrupole anisotropy
induced by superhorizon-size fluctuations of lengthscale $L$ is
$Q_L\simeq (\delta \rho/\rho)_L (LH_0 )^{-2}$. The COBE DMR has
measured a quadrupole anisotropy of
$Q_{COBE} =(4.8\pm1.5)\times 10^{-6}$ from the first year of data
and $Q_{COBE} = (2.2\pm1.1)\times 10^{-6}$ from the first two years
of data \cite{b94}.
Consequently, assuming order unity density fluctuations on scales
$L \agt L_0$,
the size of the inflated patch must be significantly
larger than the present Hubble radius, $L_0> 500H_0^{-1}$ \cite{mt,gris}.
However, the Grischuk-Zel'dovich analysis was performed for a
spatially flat ($k=0$) universe; to self-consistently exclude
an open model, it must be extended to the case of negative curvature.
We do this below and find that this constraint on $L_0$ generally becomes
even tighter when $\Omega_0 < 1$. If the Universe began from inhomogeneous
initial conditions, the
comoving size of the quasi-homogeneous patch that encompasses our
observable universe must extend to at least $500-2000$ times
the present Hubble radius. Thus, the required large
size of the inflated patch is very improbable in low-$\Omega$
inflationary models.

To relate the size of the inflated patch to the
local value of $\Omega_0$ we write the Friedmann equation as
$-K=[1-\Omega(t)]H^2(t)a^2(t)$,
where $a(t)$ is the global expansion factor and $K=+1,-1,$ or 0 is the
spatial curvature constant. Note that the global topology of the
inflationary Universe could be quite complex, with {\it e.g., } locally
Friedmann universes of both positive and negative spatial curvature
connected by wormhole throats. We will focus on the open, negatively
curved ($K=-1$) model, since it is the open model that
attracts attention as an alternative to the flat Universe on observational
grounds. Thus, we can relate the present scale $L_0$ of the homogeneous
patch to its size $L_s$ at the start of inflation,
$(1 -\Omega_0 )H_0^2 L_0^2 =  (H_s^2 L_s^2 )(1 -\Omega_s )$
(where subscript `s' denotes quantitites at the onset of inflation).
By the onset of inflation, we expect that causal microphysical
processes could have smoothed out initial inhomogeneities only on scales
up to the Hubble radius, so that $H_s L_s \sim 1$. This is also
a sufficient condition for spatial gradients to be subdominant
compared to the vacuum energy density driving accelerated expansion
\cite{gp}. Inflation was
proposed in part to allow $1-\Omega_s \sim 1$ as an initial condition,
but in any case $1-\Omega_s \leq 1$. Consequently, we expect the
present size of the homogeneous patch to satisfy
$L_0^2 \alt H_0^{-2}/(1 -\Omega_0) \equiv R^2_{\rm curv}$,
{\it i.e., } the present size of
the inflated patch is at most comparable to the present
curvature radius $R_{\rm curv}$.
If $1-\Omega_0 \ll 1$, the Universe is nearly spatially flat, and the present
curvature radius is much larger than the Hubble radius. On the other hand,
if $1-\Omega \simeq 1$, then $R_{\rm curv} \sim H^{-1}_0$, implying
$L_0 \alt H^{-1}_0$, and in particular $L_0 \ll 500 H_0^{-1}$.
This simple argument shows that the GZ effect
is only naturally suppressed in the limit $\Omega_0 \rightarrow 1$, and
that the required large size
of the homogeneous domain of our observable Universe implied by the
microwave background measurements is difficult to produce in $\Omega_0 \ll 1$
inflationary models. However, as noted above, the effect of spatial
curvature on the GZ anisotropy can be significant,
and this calculation should be done self-consistently in an open universe.

Microwave background anisotropies
in an open universe have been studied by a number of authors \cite{w83as86}.
We write the background metric of the open universe in the form
$ds^2=a^2(\eta)[d\eta^2-d\chi^2-\sinh^2(\chi)(d\theta^2+\sin^2\theta
d\phi^2)] $~,
where $\eta = \int dt/a(t)$ is conformal time, and
$\chi$ is the comoving radial distance in units of the curvature scale
({\it i.e., } the physical distance $\chi_{\rm phys} = R_{\rm curv} \chi$).
For the matter-dominated universe, the scale factor is given by
$a(\eta)=a_m \,
(\cosh \,\eta -1)$, where $a_m$ is a constant and $\eta=0$
corresponds to the initial singularity. At a given conformal time,
the density parameter is given by $\Omega (\eta )=2(\cosh\, \eta
-1)/\sinh^2\eta$. Thus, at early times, $\eta \ll 1$, the universe is
effectively
flat, $\Omega(\eta \ll 1) \simeq 1$, and at late times,
$\eta \agt 1$, it is curvature-dominated.

To describe the propagation of waves in curved space, we expand them in
terms of eigenfunctions of the Helmholtz equation
$(\nabla^2 + k^2 +1)f(\chi,\theta,\phi)=0$,
where $\nabla^2$ is the Laplace operator on the three-surface of constant
negative curvature. The solutions are of the form
$X_l(k;\chi)Y_l^m(\theta,\phi)$, where $Y_l^m$ are the spherical harmonics
and the radial eigenfunctions are given by
\begin{equation}
X_l(k;\chi ) =
(-1)^{l+1}{(k^2+1)^{l/2} \over N_l^{-1}(k)} \sinh^l\chi{d^{l+1}(\cos k\chi )
\over d(\cosh\chi )^{l+1}}  \,\, .
\label{xl}
\end{equation}
Here $N_l^k =k^2(k^2+1)...(k^2+l^2)$. The normalization is chosen
such that in the limit $\Omega \rightarrow 1$ the radial
eigenfunctions become spherical Bessel functions.
For
a perturbation of comoving wavelength $\lambda$, the
comoving wavenumber $k=2\pi/\lambda = k_{\rm phys} R_{\rm curv}$.
Using this relation and the Friedmann equation, the comoving wavenumber
corresponding to the size of the inflated patch is
$k_0 \simeq R_{\rm curv}/L_0 = 1/L_sH_s \sqrt{1-\Omega_s} \agt 1$.

Similarly,
the microwave background temperature can be expanded in spherical
harmonics on the sky, $\delta T/T = \sum a_{lm} Y_{lm}(\theta,\phi)$,
and the multipole moments of the anisotropy are then given by the
Sachs-Wolfe relation
\begin{equation}
\langle |a_l|^2 \rangle= \frac{2}{\pi}
\int |\Phi_k (\eta=0)|^2
|\tilde{\theta}_l(k)|^2 \frac{N_l(k)}{(k^2+1)^l} dk ~~,
\label{sw}
\end{equation}
where
\begin{equation}
\tilde{\theta}_l(k)={F(\eta_{ls}) \over 3}X^l_k(\eta_0-\eta_{ls})+2
\int_{ls}^{\eta_0} {dF \over d\eta} X^l_k(\eta_0-\eta )d\eta \,\, ,
\label{th}
\end{equation}
$\eta_{ls}$ denotes the epoch of last scattering,
and the gravitational potential fluctuation satisfies
$\Phi_k(\eta) = \Phi_k(\eta=0) F(\eta)$, with (ignoring the decaying
mode)\cite{m92}
\begin{equation}
F(\eta) =5{\sinh^2\eta -3\eta\sinh\eta+4\cosh\eta -4 \over (\cosh\eta
-1)^3} ~.
\label{feta}
\end{equation}
Note that $F(\eta)=1 $ for $\Omega_0=1$; in an open universe,
$F(\eta) \simeq 1$ for $\eta \alt 1$ and decays as $1/a(\eta)$
for $\eta \agt 1$.
Eqs.\ (\ref{xl}) - (\ref{feta})
allow one to estimate the anisotropy due to superhorizon-size
perturbations, with wavelengths $\lambda \gg H^{-1}_0$.
The potential $\Phi$ is a gauge-invariant measure of the spatial curvature
perturbation, related to the density fluctuation by the relativistic
curved-space analogue of the Poisson equation \cite{b80}.
For perturbations on scales larger than the Hubble radius, $k\eta \alt 1$,
it satisfies $\Phi_k \simeq -\delta_k/2 \simeq$
constant, where $\delta$
is a gauge-invariant measure of the density perturbation amplitude,
equal to the density fluctuation in the longitudinal (conformal Newtonian)
gauge \cite{m92}. For such long wavelengths, the dominant anisotropy
is generally the quadrupole $l=2$ (for some values of
$\Omega_0$, the quadrupole is accidentally suppressed, and the main
contribution would be the $l=3$ octupole moment, as we discuss below).
The quadrupole anisotropy due to such superhorizon-scale modes is thus
\begin{equation}
\langle |a_2|^2\rangle \simeq {1 \over 2\pi} \int_0^{k_0} {k^2 +4\over k^2 +1}
|\tilde{\theta}_2(k)|^2  \langle |\delta_k|^2\rangle k^2 dk  \,\, .
\label{cl2}
\end{equation}

Fig.1 shows the mode contribution $|\tilde{\theta}_2(k)|^2$ to the
quadrupole, for several values of $\Omega_0$, for modes outside the
present Hubble radius. The suppression at $\Omega_0 = 0.4$ arises
from a near cancellation of the line-of-sight
contribution (the second term on the RHS of eq.\ (\ref{th})) with the last
scattering term (see below).
We emphasize that for modes outside the scale of the homogeneous patch,
$k \leq k_0$, the pre-inflation perturbation amplitude $\delta_k$ is
preserved and expected to be of order unity.
To study the implications of this result, we consider
two limits: $\Omega_0$ close to unity ($1-\Omega_0 \ll 1$) and
low-density models with $1-\Omega_0 \sim 1$.

\psfig{file=fig1.eps,height=12cm,width=15cm}

{\footnotesize{ Fig.\ 1: The quadrupole mode contribution for superhorizon-size
perturbations,
$|\tilde{\theta}_2|$, as a function of $k$ for $\Omega_0 = 0.1, 0.4,$ and 0.7.
Each curve ends at the value of $k$ corresponding to modes just entering the
present Hubble radius, $k\eta_0 =1$.}}
\vspace{0.8cm}

{\bf $\Omega_0$ close to 1}: Using the relation $\cosh \eta - 1 = 2(1-\Omega)/
\Omega$, the limit $\Omega_0 \rightarrow
1$ corresponds to taking
$\eta^2_0 \simeq 4(1-\Omega_0) \rightarrow 0$. Taking this limit
in Eq.\ (\ref{xl}) while keeping $k_{\rm phys}$ fixed, we find
$X_2(k;\eta\rightarrow 0)
\simeq (1+k^2)\eta^2/15  \simeq
4\, (1+k^2)(1-\Omega_0)/15$. In this limit,
the line-of-sight integral in Eq.\ (\ref{th}) becomes
$\int (dF/d\eta) X^2_kd\eta
\simeq -(1+k^2)\eta_0^4/630$, which can be neglected
compared to the last scattering term. As a result, the
quadrupole arising from modes with $k\eta_0 \alt 1$ can be expressed as
\begin{equation}
\langle |a_2|^2\rangle \simeq
{8 \over 225 \pi}{(1-\Omega)^2 \over 9} \int_0^{k_0} dk ~ k^2
(k^2+1) (k^2+4) \langle|\delta_k|^2\rangle ~~.
\label{C2}
\end{equation}
The usual flat-space result can be recovered from Eq. (\ref{C2})
by taking the limit $k \gg 1$ and keeping $k_{\rm phys}$
fixed in the relation $k_{\rm phys} = kH_0\sqrt{1-\Omega_0}$.
Eq.\ (\ref{C2}) can be used to constrain $\Omega_0$ with any given
pre-inflation power spectrum $\langle|\delta_k|^2\rangle$ on
scales $k \leq k_0$. A plausible assumption is that
$\langle|\delta_k|^2\rangle \sim k^n$ with $n \geq 0$, {\it i.e., } random
Poisson
fluctuations (or less). For example,
such a spectrum would arise if one imagines that prior to
inflation the universe consisted of uncorrelated, quasi-homogeneous regions of
size $k_0^{-1}$. However, quantitatively the result does not depend strongly
upon the shape of the power spectrum. With the assumption of no fine tuning
prior to inflation, {\it i.e., } $\langle|\delta(k_0)|^2\rangle
\simeq 1$, and since in inflationary models $k_0 \agt 1$, the COBE
measurement of the quadrupole moment translates
eq.\ (\ref{C2}) into the constraint
\begin{equation}
\Omega_0 > 1- a_2({\rm COBE}) \simeq 1 - 10^{-6} ~~.
\end{equation}
Thus, {\it if} an epoch of inflationary expansion
was responsible for the homogeneity of our observable Universe, the density
parameter $\Omega_0$ cannot differ from 1 by more than one part in
$Q^{-1} \sim 10^6$.

{\bf Low $\Omega_0$}:
We now consider the case of low $\Omega_0$ and estimate the scale out
to which
the Universe must be homogeneous in light of the COBE results, independent
of considerations of inflation, namely we allow $k_0 \ll 1$. If $\Omega_0$ is
not very close to 0.4 or
1, Fig. 1 shows that
$\tilde{\theta}_2(k)$ is nearly independent of $k$ for small $k$, and we can
set $\tilde{\theta}_2(k) \simeq \tilde{\theta}_2(0)$ to good approximation
for $k <1$. (This is very different from the spatially flat model, where
$|\tilde{\theta}_2(k)|
=j_2(2k)$ and goes to zero as $k^2$ at small $k$).
The zero-mode contribution
$|\tilde{\theta}_2(0)|$ as a function of $\Omega_0$ is shown in Fig. 2.
In this case the quadrupole becomes:
\begin{equation}
\langle |a_2|^2\rangle \simeq |\tilde{\theta}_2(0)|^2 {1 \over 2\pi}
\int_0^{k_0} {k^2 +4\over k^2 +1} \langle|\delta_k|^2\rangle k^2 dk ~~.
\label{lowden}
\end{equation}
Again conservatively assuming an
initial spectrum that falls at least as white noise ($n \geq 0$),
eq.\ (\ref{lowden}) yields a lower bound on the scale $k_0^{-1} \sim L_0H_0
\sqrt{1-\Omega_0}$
over which the Universe must be homogeneous if $\Omega_0 \alt 1$,
\begin{equation}
k_0^{-1} >
\left(\frac{|\tilde{\theta}_2(0)|}{a_{2,COBE}}\right)^{2/3}
\simeq 10^4 |\tilde{\theta}_2(0)|^{2/3} ~~.
\label{limit}
\end{equation}
Using Fig. 2, we see that eq.\ (\ref{limit}) implies
that the Universe
must to be homogeneous over scales  $k_0^{-1}\agt 2000 $ for
$\Omega_0 \alt 0.1$
and over scales $k_0^{-1}\agt 500 $ for $\Omega \simeq 0.5 - 0.8$.
In inflation models these bounds on $k_0 \ll 1$ require superhorizon-sized
correlations prior to inflation. Note that for a given constant value of
$|\delta_k|^2$ the quadrupole anisotropy for $k_0 \ll 1$ scales as
$\sim k_0^{7/2}$ for $\Omega =1$ and only as $\sim k_0^{3/2}$ for $\Omega \ll
1$.


\psfig{file=fig2.eps,height=12cm,width=15cm}

{\footnotesize{Fig.\ 2: $|\tilde{\theta}_2|$ at $k=0$ as a function of
$\Omega_0$.}}
\vspace{0.8cm}

{\bf $\Omega_0 \simeq 0.4$}:
As Figs. 1 and 2 illustrate, the quadrupole due to long wavelength modes is
suppressed not only at $\Omega_0 \rightarrow 1$ but also accidentally for
$\Omega_0 \sim 0.4$, due to cancellation between the last scattering term and
the line-of-sight integral. (The positive last scattering term
dominates at $\Omega_0 \rightarrow 1$, while the negative line-of-sight
term dominates at $\Omega_0 \rightarrow 0$.) As $\Omega_0$ is varied
over a small interval around 0.4, the wavenumber where the two terms
cancel varies over the interval ($0,\eta_0^{-1}$). While interesting,
this suppression cannot make inflation and
low-$\Omega_0$ compatible, for in this case the
contribution
to the octupole ($l=3$) mode will be dominant and lead
to similarly severe constraints on $L_0$.

We have arrived at two results of significance for inflation and
open universe models.
(1) Inflation can produce a
homogeneous patch encompassing the observable Universe (the present
Hubble volume)
and be consistent with the microwave background observations
only if the present density parameter $\Omega_0$ differs from unity
by no more than 1 part in $Q^{-1}_{COBE} \sim 10^6$.
(2) On the other hand, if $\Omega_0$ is significantly below 1,
the Universe must be homogeneous on scales $k_0^{-1}>(500-2000)$.
If this is the case, inflation does not by itself solve the horizon problem.
Indeed, if we assume that
the distribution of quasi-homogeneous regions
satisfies Poisson statistics, the probability of
finding one such region per volume $k_0^{-3}$ in curvature
units
is $P\simeq k_0^{-3}\times \exp(-k_0^{-3})$, which
is negligibly small for the $k_0^{-1}$ values above.
If it turns out that the universe is open,
$\Omega_0 <1$, this implies that our Hubble volume
occupies a very special place in the space of initial conditions,
which is precisely the condition inflation was meant to alleviate.
One might then prefer to seek a solution of the horizon problem
in quantum gravitational effects at the Planck era.

We thank R. Caldwell, M. Kamionkowski, A. Liddle,
and A. Stebbins for useful discussions.
This work was supported by the National Science Foundation under
Grant No.\ PHY89-04035, and by the DOE and NASA grant NAGW-2381
at Fermilab. AK was supported by the NRC and acknowledges the hospitality
of Theoretical Astrophysics Group at Fermilab.

\end{document}